\documentclass[trackchange,twocolumn]{aastex701}

\usepackage{adjustbox}
\usepackage{amsmath}

\newcommand{\oi}{\hbox{[O$\,${\scriptsize I}]}}
\newcommand{\oii}{\hbox{[O$\,${\scriptsize II}]}}
\newcommand{\oiii}{\hbox{[O$\,${\scriptsize III}]}}

\newcommand{\feii}{\hbox{[Fe$\,${\scriptsize II}]}}

\newcommand{\sii}{\hbox{[S$\,${\scriptsize II}]}}

\newcommand{\hei}{\hbox{He$\,${\scriptsize I}}}
\newcommand{\nii}{\hbox{[N$\,${\scriptsize II}]}}

\newcommand{\kms}{km\,s$^{-1}$}

\newcommand\astropy{\texttt{Astropy}}
\newcommand\numpy{\texttt{NumPy}}
\newcommand\scipy{\texttt{SciPy}}
\newcommand\pyneb{\texttt{PyNeb}}
\newcommand\photutils{\texttt{Photutils}}
\newcommand\matplotlib{\texttt{matplotlib}}

\newcommand\badass{\texttt{BADASS}}
\newcommand\cwitools{\texttt{cwitools}}
\newcommand\ppxf{\texttt{pPXF}}

\shorttitle{KCWI Observations of NGC 1275}
\shortauthors{Dan, et al.}
\submitjournal{ApJ}

\begin{document}

\title{KCWI Discovery of a Spatially Resolved Kpc-Scale Ionized Outflow in NGC 1275} 

\author[orcid=0000-0001-5894-4651]{Kylie Yui Dan}
\affiliation{Department of Astronomy, University of Maryland, College Park, MD 20742, USA}
\affiliation{Hiroshima Astrophysical Science Center, Hiroshima University, 1-3-1 Kagamiyama, Higashi-Hiroshima, Hiroshima 739-8526, Japan}
\email[show]{kydan@umd.edu}

\author[orcid=0000-0003-4268-0393]{Hanae Inami}
\affiliation{Hiroshima Astrophysical Science Center, Hiroshima University, 1-3-1 Kagamiyama, Higashi-Hiroshima, Hiroshima 739-8526, Japan}
\email{hanae@hiroshima-u.ac.jp}

\author[orcid=0000-0002-6562-8654]{Thomas Bohn}
\affiliation{Center for Cosmic Evolution Research, Ehime University, Bunkyo-cho 2-5, Matsuyama, Ehime 790-8577, Japan}
\email{bohn.thomas_carl.gb@ehime-u.ac.jp}

\author[orcid=0000-0002-3158-6820]{Sylvain Veilleux}
\affiliation{Department of Astronomy, University of Maryland, College Park, MD 20742, USA}
\affiliation{Joint Space-Science Institute, Department of Astronomy, University of Maryland, College Park, MD 20742, USA}
\email{veilleux@umd.edu}

\author[orcid=0000-0002-1912-0024]{Vivian U}
\affiliation{IPAC, California Institute of Technology, 1200 E. California Blvd., Pasadena, CA 91125, USA}
\email{vivianu@ipac.caltech.edu}

\author[orcid=0000-0002-1158-6372]{Tianmu Gao}
\affiliation{Research School of Astronomy and Astrophysics, Australian National University, Weston Creek, ACT 2611, Australia}
\affiliation{ARC Centre of Excellence for All Sky Astrophysics in 3 Dimensions (ASTRO 3D), Australia}
\email{tianmu.gao@anu.edu.au}

\author[orcid=0000-0002-6650-3757]{Justin Kader}
\affiliation{IPAC, California Institute of Technology, 1200 E. California Blvd., Pasadena, CA 91125, USA}
\email{kaderj@uci.edu}

\author[orcid=0000-0003-2064-4105]{Rosalie McGurk}
\affiliation{W. M. Keck Observatory, Kamuela, HI, USA}
\email{rmcgurk@keck.hawaii.edu}


\begin{abstract}
We present new {\em Keck Cosmic Web Imager} observations of the central few kiloparsecs of NGC 1275, the brightest cluster galaxy of the Perseus Cluster. These integral field spectroscopic data reveal a warm-ionized outflow traced by H$\beta$ and the \oiii\ doublet extending out to $\sim2.5$ kpc from the nucleus. The warm-ionized outflow has an \oiii-derived $v_{50}$ of up to $\sim570$ \kms, $w_{80}$ of up to $3780$ \kms, and H$\beta$-derived outflowing mass of $(2.7 \pm 0.3)\times 10^6$ M$_\odot$. Our H$\beta$-derived warm-ionized outflowing mass rate of 2.7$\pm$0.7 M$_\odot$ yr$^{-1}$ is comparable to the estimated cold molecular disk accretion rate of 1$-$10 M$_\odot$ yr$^{-1}$, which could be a sign of self-regulation between the pc-scale cold gas feeding the active galactic nucleus (AGN) and the warm wind accelerated out to kpc scales. In the host galaxy, we detect an enhancement in the \oiii\ / H$\beta$ ratio in the direction of the receding jet, which may imply jet interaction with the host interstellar medium. The outflow component also shows a clear enhancement of \oiii\ / H$\beta$ that positively correlates with $w_{80}$ and $v_{50}$, indicative of AGN influence and/or fast shocks. Both the outflow and host galaxy \oiii\ / H$\beta$ ratio decrease with increasing distance from the center, showing that the influence of the AGN is confined to the central $\sim2$ kpc.  These results are the first resolved measurements of a kpc-scale ionized wind in NGC 1275 and provide a view into how both jets and winds contribute to the feedback cycles in complex cool core cluster systems. 

\end{abstract}


\section{Introduction} 
Brightest cluster galaxies (BCGs), which reside at the centers of massive galaxy clusters, have long been associated with the “cooling flow problem,” where the surrounding intracluster medium (ICM) is expected to cool efficiently via X-ray emission, leading to inferred mass accretion rates of $\sim10-1000$ $M_\odot$ yr$^{-1}$ \citep{fabian1994}. However, the observed lack of cooling gas and star formation relative to these predictions indicates that radiative cooling must be offset by an additional heating mechanism. Observations now strongly support AGN feedback as the dominant heating source, with relativistic jets and outflows depositing mechanical energy into the ICM and regulating cooling \citep[e.g.,][]{veilleux2005, mcnamara2007, fabian2012}. 

NGC 1275 \citep[z = 0.01756;][]{hitomi2016}, the BCG of the Perseus Cluster, has been the primary laboratory for studying these processes. It is well-known for its spectacular optical filaments that extend to 10s of kpc into the cluster and are thought to be dragged out from the center of the galaxy by radio bubbles rising buoyantly in the hot intracluster gas \citep{Conselice2001, Fabian2008, Vigneron2024, Rhea2025}. These radio bubbles are driven by the central AGN (also referred to as 3C 84), which exhibits powerful, parsec-scale jets that expand into the dense interstellar medium (ISM), inflating the large-scale radio bubbles that show up as distinct X-ray cavities in Chandra imaging \citep{Fabian2006, Sanders2007}. A small, secondary galaxy referred to as the ``High Velocity System" (HVS) is thought to be infalling at a projected velocity offset of $\sim3000$
 \kms \citep[e.g.;][]{lynds1970, rubin1977, Conselice2001, Yu2015, Rhea2025_HVS}. 

In contrast to the extended filamentary structures that have been mapped in great detail, the inner few kpc represent the crucial interface where energy is first injected from the AGN into the host galaxy’s ISM. 
Previous observations of the central regions of NGC 1275 show various inflows and outflows. Gemini near-infrared integral field spectrograph (NIFS) observations of H$_2$ 1$-$0 S(1) in the inner 100 pc reveal a hot molecular inflow up to 500 \kms\ feeding a 50 pc-scale hot molecular disk \citep{scharwachter2013}, while similar Gemini NIFS observations of the inner 900 pc find outflows traced by H$_2$ 1$-$0 S(1) and \feii\ 1.2570 $\mu$m with velocity offsets of $\pm 150-200$ \kms\ and 80th-percentile widths up to 2000 \kms\ extending up to 500 pc from the center \citep{riffel2020}. Atacama Large Millimeter Array (ALMA) observations of HCN(3$-$2) and HCO+(3$-$2) detect outflowing cold molecular clouds with velocities $\sim -300-600$ \kms\ as well as a 100 pc-scale cold molecular disk \citep{nagai2019}. \citet{oosterloo2024} reprocessed the ALMA observations from \citet{nagai2019} to show how the kpc-scale filaments feed the cold molecular disk with complicated velocity structures. Notably lacking is the detection of an ionized gas wind, often found in AGN-dominated systems. Even deep Chandra High-Energy Transmission Grating observations fail to detect a photoionized wind \citep{reynolds2021}. Characterizing the ionized gas phase is particularly crucial for understanding the energetics of these outflows, as the ionized component typically exhibits the most extreme kinematics, often surpassing the velocities of the neutral and molecular phases \citep[e.g.,][]{veilleux2005, rupke2013, veilleux2020}. 

In this work, we present a detailed view into the central 10 kpc of NGC 1275 using the optical ionized gas tracers H$\beta$ and \oiii\ 5007 \AA, resolving for the first time a kpc-scale warm ionized ($10^4$ K) outflow. In Section \ref{sec:reduction}, we go over the observations and data reduction process. We describe our analysis methods in Section \ref{sec:analysis} and present the results in Section \ref{sec:results}. In Section \ref{sec:discussion}, we discuss the \oiii\ / H$\beta$ ratio as well the mass, energetics, and influence of the outflow. We summarize our conclusions in Section \ref{sec:conclusion}. We adopt a cosmology of $H_0 = 70$ \kms Mpc$^{-1}$, $\Omega_M = 0.3$, and $\Omega_L = 0.7$, which implies a scale of 0.357 kpc arcsec$^{-1}$ and a luminosity distance of 76.2 Mpc.

\begin{figure}
    \centering
    \includegraphics[width=0.9\columnwidth]{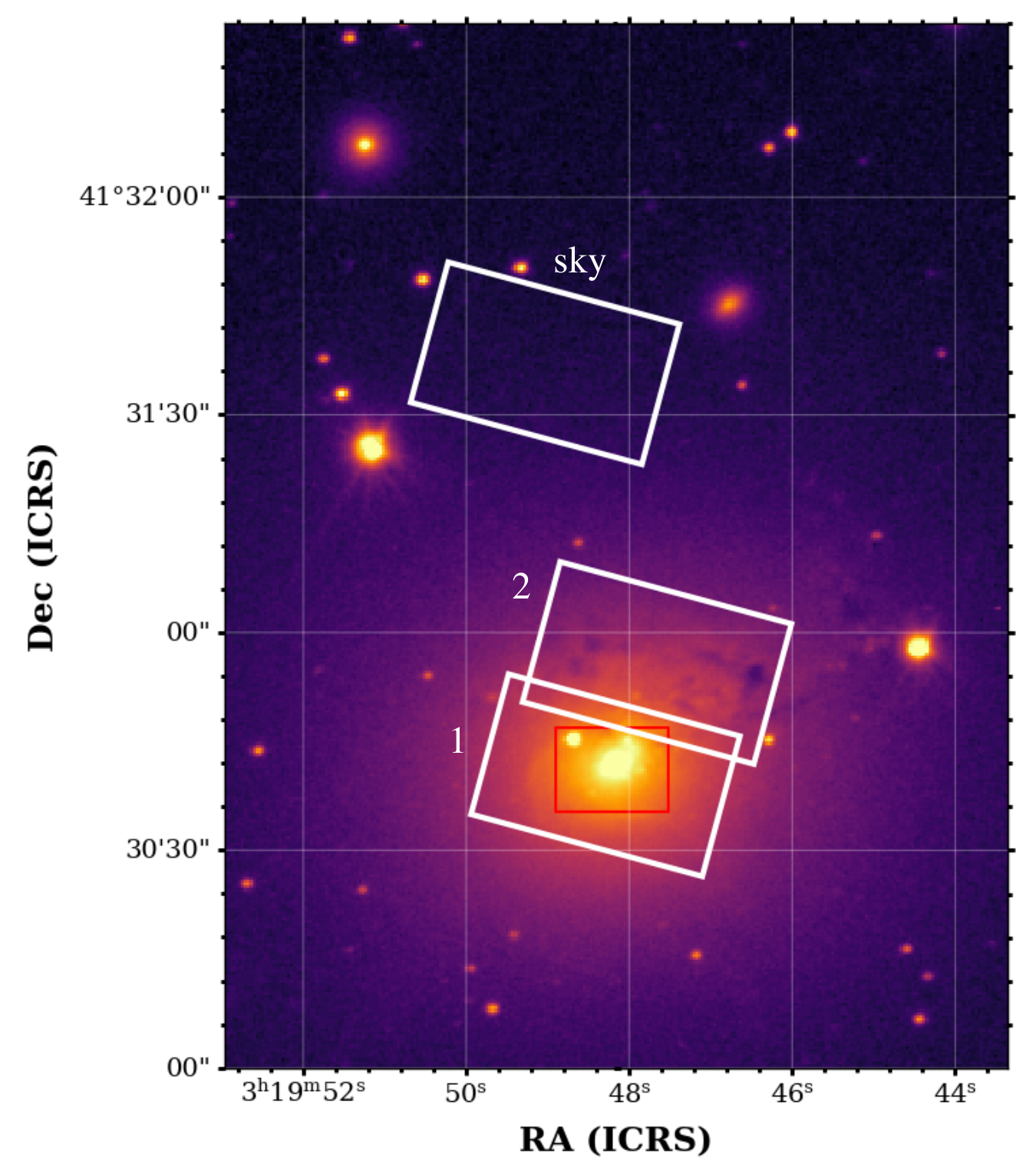}
    \caption{SDSS image of NGC 1275 where white rectangles labeled 1 and 2 represent the two science pointings separated by $\sim 17\arcsec$, and the rectangle labeled ``sky'' marks the background pointing. Pointing 1 focuses on the central region of NGC 1275 while pointing 2 captures part of the extended optical filaments and HVS. The red box within pointing 1 is the same as in Figure \ref{fig:maps}. 
    \label{fig:pointings}}
\end{figure}

\section{Observations and Data Reduction}
\label{sec:reduction}

NGC 1275 was observed with the W. M. Keck Observatory\footnote{
The data presented herein were obtained at Keck Observatory, which is a private 501(c)3 non-profit organization operated as a scientific partnership among the California Institute of Technology, the University of California, and the National Aeronautics and Space Administration. The Observatory was made possible by the generous financial support of the W. M. Keck Foundation.} on UT 2021-09-14 using the BH3-L mode (high spectral resolution grating with the large slicer) of the Keck Cosmic Web Imager \citep[KCWI;][PID 2021B-K318, PI Lewis, ``Studying Multiphase Feedback in LIRGs with KCWI and JWST"]{kcwi2018} with a rest wavelength coverage of 4764\AA\ to 5261\AA, field of view of 33'' $\times$ 20'', spatial resolution of 1.39'', and spectral resolution of $\sim$4500. The sky conditions were clear with an estimated seeing of 0.63''. Two pointings were taken, one centered on the nucleus of NGC 1275 with a combined exposure time of 540 s and another offset by $\sim 17\arcsec$ north-northwest to capture both the extended filamentary structure and the infalling HVS with a combined exposure time of 2040 s. A sky background pointing, twilight flats, and standard star measurements were also taken. The pointings are shown in Figure \ref{fig:pointings}. The data used in this paper can be found in the Keck Observatory Archive (KOA; DOI:\href{https://www.ipac.caltech.edu/doi/10.26135/KOA8}{10.26135/KOA8}).

We reduced these data with the KCWI Data Reduction Pipeline (DRP) v1.1 \citep{kcwidrp2023}, publicly available as a Python package (\href{https://kcwi-drp.readthedocs.io/en/latest/}{kcwi-drp}). The DRP has 6 main steps:
\begin{enumerate}
    \item Bias calibration
    \item Continuum bars calibration
    \item Wavelength calibration via Thorium-Argon arc lamp observations
    \item Flat-field calibration
    \item Flux calibration via standard star observations
    \item Sky background subtraction
\end{enumerate}
We first reduced the sky background pointing to produce a master sky file, then reran the DRP a second time to reduce the science images. To combine the two reduced science pointings, we utilized the Python package \cwitools\ to first crop the edges of the cubes to remove bad spaxels and then coadd the two pointings into a single $124 \times 140$ spaxel cube.

\section{Data Analysis}
\label{sec:analysis}

To fit the H$\beta$ and \oiii\ doublet emission lines, we use the Bayesian AGN Decomposition Analysis for Sloan Digital Sky Survey (SDSS) Spectra \citep[\badass\ v11.0.0;][]{sexton2021} code. Through Markov Chain Monte Carlo routines, \badass\ performs simultaneous multicomponent fits to emission-line spectra. To determine an appropriate continuum fit, we compare fits between two different models: a simple AGN power law and a more complex host galaxy with a stellar line-of-sight velocity distribution (LOSVD) via the penalized pixel fitting method (\ppxf; \citet{Cappellari2004}; templates from Indo-U.S. Library of Coudé Feed Stellar Spectra, \citet{Valdes2004}). We fit the entire cube with both continuum options and utilize the Bayesian information criterion (BIC) test to establish that the LOSVD continuum model is statistically better for nearly all spaxels. For continuity, we thus utilize the LOSVD continuum for all spaxels. 

Emission lines are modeled using up to three Gaussian profiles, with the number of Gaussians selected based on line-testing options implemented within \badass. For the test options, we used a combination of the self-titled BADASS statistic $> 0.95$, which generates a confidence between 0 and 1 of the Monte-Carlo resampled maximum likelihood values of the simple and complex models and justifies the additional component for a confidence $> 0.95$, and amplitude-over-noise (AON) $> 3$, which determines if the lines that comprise the best model have an AON above 3. Each additional Gaussian component is restricted to have a larger width than the previous one (e.g. component 1 has the smallest width and component 3 has the largest). The width and velocity offsets of each component are tied between H$\beta$ and the \oiii\ doublet, and the amplitudes are set to freely vary. 

For further analysis, we separate the components into a narrow component, representing the low-velocity galaxy emission, and an ``outflow" component comprised of the additional one or two broader fitted component(s). When two additional fitted components are included, their fits are combined and we report velocity offsets and widths of the combined profiles as the ``outflow" component. We utilize this process for all spaxels in the KCWI cube, and afterwards applied a signal-to-noise ratio (SNR) cut of 3 when creating the parameter maps, where the SNR is calculated within \badass\ as the maximum value of the line model above the mean value of the noise. 

\begin{figure*}
    \centering
    \includegraphics[width=0.8\textwidth]{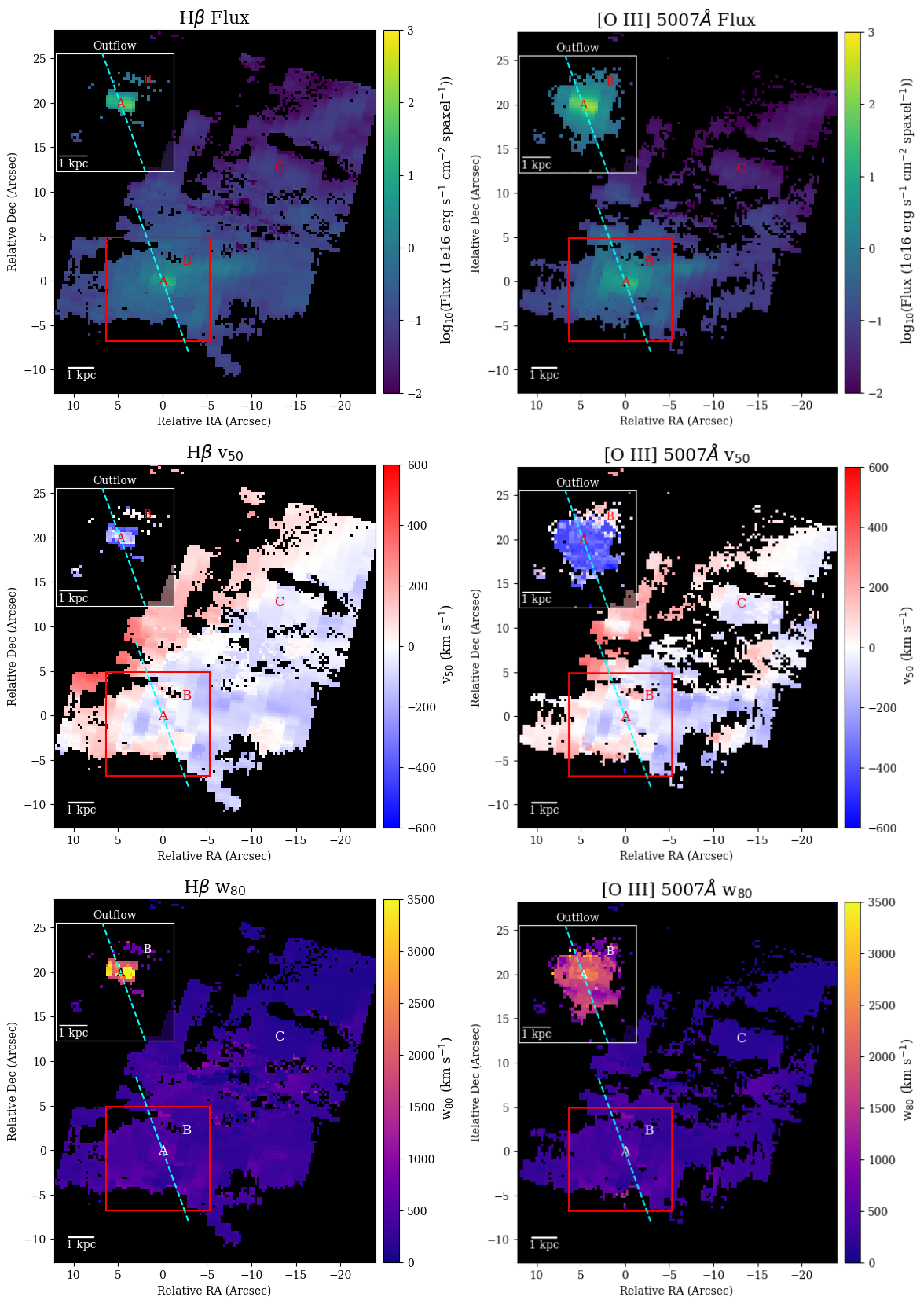}
    \caption{Left column (top to bottom): H$\beta$ flux, $v_{50}$, and $w_{80}$ maps of NGC 1275. Right column (top to bottom): \oiii\ 5007\AA\ flux, $v_{50}$, and $w_{80}$ maps. In all subplots, the main image represents the host galaxy component with the outflow component displayed in the inset. The red boxes highlight the extent of the inset within the host galaxy. Fits of the marked spaxels A, B, and C are shown in Figure \ref{fig:spec} where A marks the central peak emission, B shows a spaxel with clear redshifted outflow emission, and C displays a spaxel with emission lines from both NGC 1275 and the HVS. In all panels, north is up and east to the left. The dashed cyan line shows the radio jet position angle measured within the central 20\arcsec.
    \label{fig:maps}}
\end{figure*}

\section{Results}
\label{sec:results}
\begin{figure}
    \centering
    \includegraphics[width=\columnwidth]{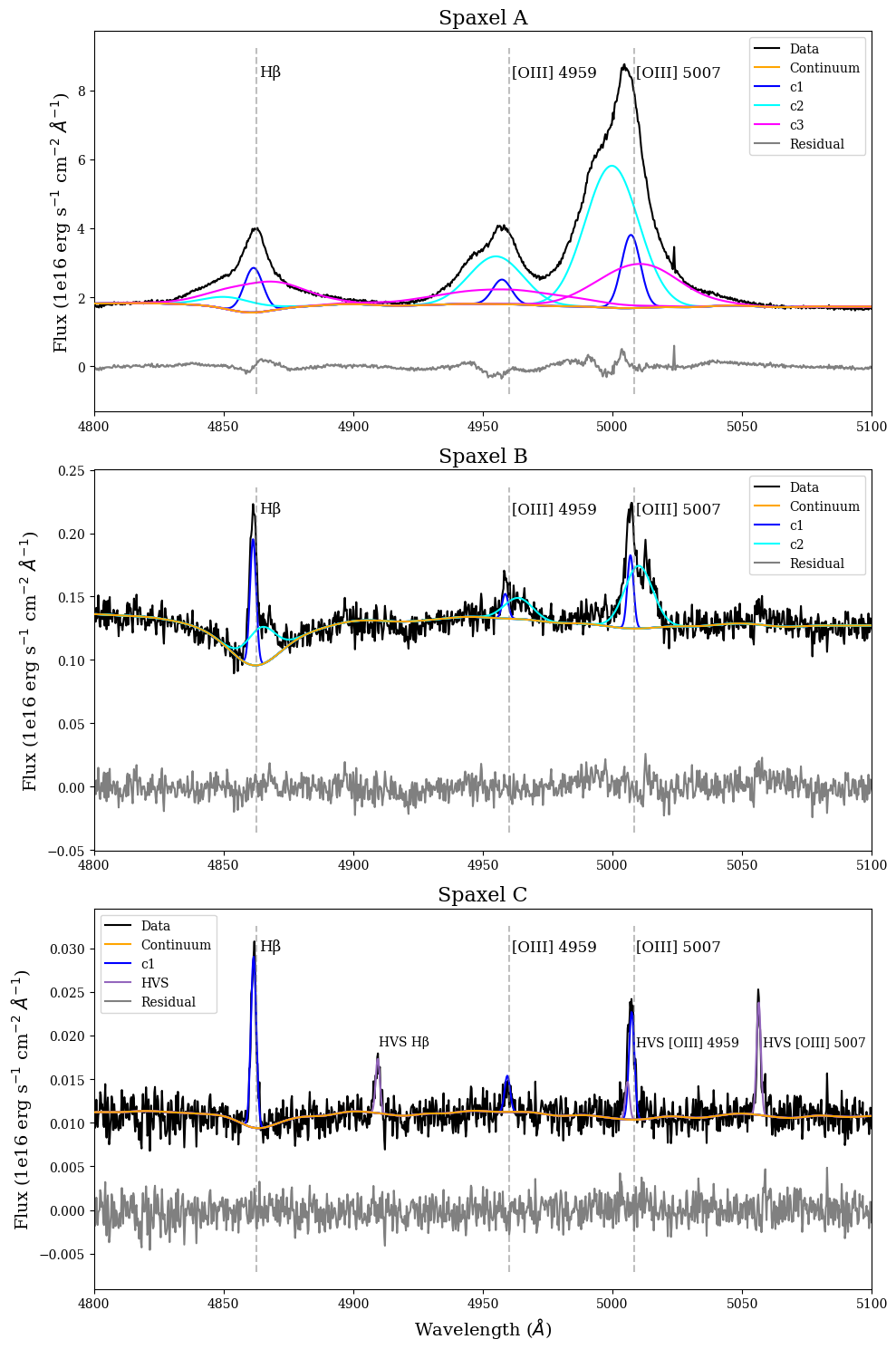}
    \caption{Example of single spaxel fits. Data is shown in black, continuum fit in yellow, narrow (host galaxy) component 1 in blue, broad outflow components 2 and 3 (if available) in teal and magenta, HVS components (if available) in purple, and residuals in gray. Top: Spectrum of the central spaxel of the nucleus of NGC 1275. Complex emission line profiles require 3 components for each. Middle: Spaxel slightly to the north-west of the nucleus, representing a slim area with redshifted outflow emission. Bottom: Spaxel to the far north-west, showing emission lines from both the main galaxy and the HVS.
    \label{fig:spec}}
\end{figure}

Maps of the flux, median velocity ($v_{50}$) and 80-percentile line widths ($w_{80} \equiv | v_{90} - v_{10} |$, where $v_{90}$ and $v_{10}$ are respectively the 90- and 10-percentile velocities) for both H$\beta$ and \oiii\ 5007 \AA\ are shown in Figure \ref{fig:maps}, where the main images represent the host galaxy narrow Gaussian component and the smaller insets represent the combined c2 and c3 ``outflow" component. Red boxes are overplotted on the main images to show the location of the outflow in the host galaxy. Example fits for three individual spaxels (labeled A, B, and C on the maps) are shown in Figure \ref{fig:spec}. Spaxel A marks the center of the nucleus of NGC 1275 and contains broad, complex emission line profiles requiring three Gaussian components per line. Spaxel B shows the small area with solely redshifted outflow emission, and spaxel C displays lines from both the main galaxy and HVS. 

\begin{figure*}
    \centering
    \includegraphics[width=0.9\textwidth]{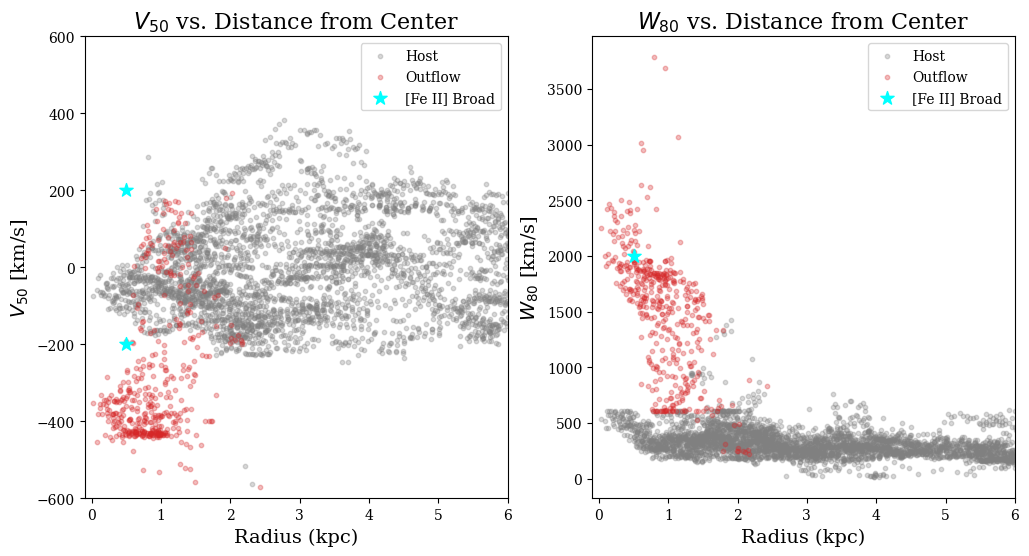}
    \caption{Left: \oiii\ $v_{50}$ vs distance from the center. Gray points represent the host galaxy component while red points show the outflow component. Right: \oiii\ $w_{80}$ vs distance from the center with the same points as the left panel. In both panels, the 500 pc scale \feii\ 1.2570 $\mu$m outflow from \citet{riffel2020} probing neutral hydrogen is shown as cyan stars. Both panels are zoomed into the central 6 kpc, but the host data extends to $\sim11.5$ kpc.
    \label{fig:pv}}
\end{figure*}

The narrow, host galaxy emission is more significantly detected in H$\beta$ than \oiii\ 5007 \AA\, and the initial few kpc of some of the well-known filaments can also be seen in the maps. Although our data do not have the spatial scale to cover all filaments, there are many other recent studies that analyze them in great detail \citep[e.g.;][]{Vigneron2024, Rhea2025}. The host galaxy Gaussian profiles are narrow ($w_{80} \lesssim 700$ \kms) and relatively low velocity ($-250 < v_{50} < 400$ \kms). In Figure \ref{fig:pv}, we show the $v_{50}$ and $w_{80}$ values of \oiii\ plotted vs radius from spaxel A, where the gray points represent the host narrow component. 
The host galaxy $v_{50}$ values vary across the maps with no clear rotational gradient, as seen in previous kpc-scale studies \citep[][]{salome2006}, although a disk has been seen in this system at tens of pc scales \citep[][]{scharwachter2013}.

In contrast to the narrow, host galaxy emission, the broad, outflow emission is more significantly detected by \oiii\ 5007 \AA. The combined outflow component is mostly blueshifted, centered around the nucleus of NGC 1275 and extending out to $\sim2.5$ kpc, with a small portion of solely redshifted emission to the northwest of the nucleus. The outflow kinematically reaches faster projected velocities than the disk ($|v_{50}| \leq 570$ \kms) and is much broader ($w_{80} \leq 3780$ \kms), marked by the red points in Figure \ref{fig:pv}. Although the $v_{50}$ values do not seem to show a trend with distance from the center, we do see a decrease in $w_{80}$ from several thousand \kms\ close to the nucleus down to several hundred \kms\ at the outflow edges several kpc from the center. For comparison, the values of the 500 pc-scale \feii\ 1.2570 $\mu$m outflow from \citet{riffel2020} are shown as cyan stars. 

\begin{figure}
    \centering
    \includegraphics[width=\columnwidth]{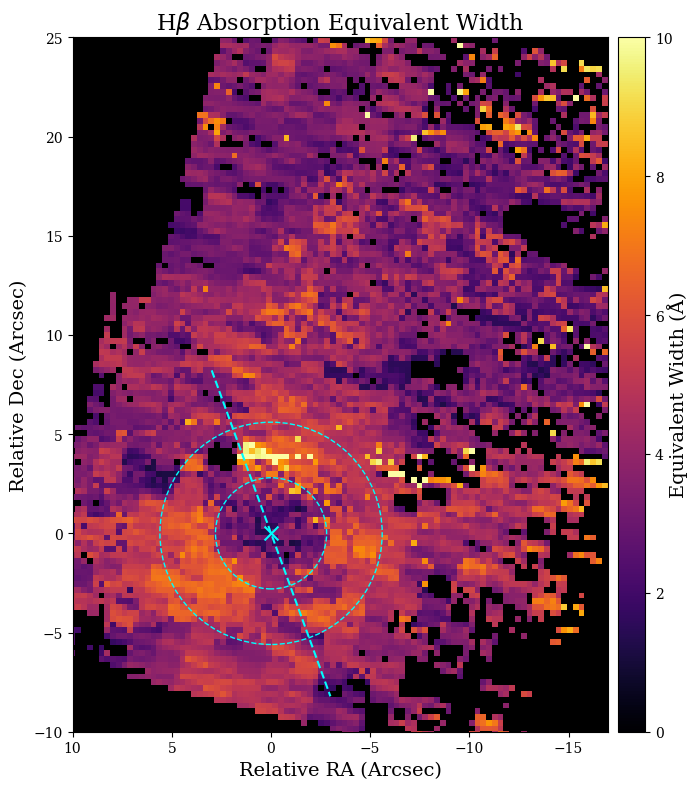}
    \caption{Map of the H$\beta$ absorption equivalent width, masking spaxels with SNR $<$ 10. Circles represent 1 and 2 kpc radii from the center, which is marked by a cross. The dashed line shows the jet angle. North is up and east to the left. 
    \label{fig:losvd_ew}}
\end{figure}

We also fit the H$\beta$ absorption feature from the LOSVD continuum with a single Gaussian, and we display a map of the equivalent width of this feature in Figure \ref{fig:losvd_ew}. The central 1 kpc shows reduced EW, indicative of AGN contamination. There is some slight enhancement along the optical filaments, but the map is too noisy to make any definite conclusions. 

\section{Discussion}
\label{sec:discussion}

We have resolved a blue- and redshifted warm ionized outflow in the nuclear region of NGC 1275 with a flux-weighted median projected velocity of $\sim 420$ \kms\ and flux-weighted 80th-percentile width of $\sim 2370$ \kms. These findings represent the first resolved measurements of the warm ionized wind in this system. The warm ionized outflow line widths are similar to the broad line region (BLR) \hei\ and Pa$\beta$ line widths \citep{onori2017}, as well as the \feii-derived outflow line widths \citep{riffel2020}. The velocities are $2-3\times$ higher than the \feii-derived values, but are remarkably consistent with the kinematics of the unresolved cold molecular absorption seen on pc scales in HCN(3$-$2) and HCO+(3$-$2) \citep{nagai2019}. The cold molecular outflow is solely observed in blueshifted absorption and attributed to dense clouds accelerated by the jet or a wind driven by AGN radiation pressure. Our detection of both blue- and redshifted emission paints the picture of a more classical biconical or spherical outflow, similarly determined by \citet{riffel2020} in \feii. 

\subsection{[O{\scriptsize III}] 5007 \AA\ / H$\beta$ Ratio}
\begin{figure}
    \centering
    \includegraphics[width=\columnwidth]{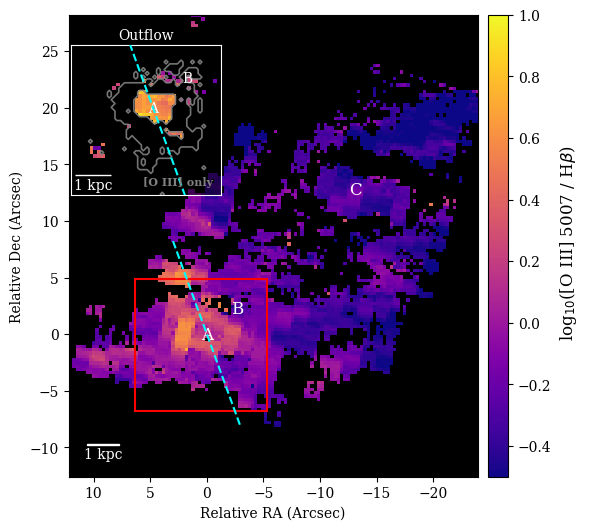}
    \caption{Map of log$_{10}$(\oiii\ 5007 \AA / H$\beta$) where the main image represents the host galaxy component with the outflow component displayed in the inset. North is up and east to the left. The red box highlights the extent of the inset within the host galaxy, and fits of the marked spaxels A, B, and C are shown in Figure \ref{fig:spec}. The dashed cyan line marks the jet direction. In the inset, the gray contours represent spaxels where an outflow component of \oiii\ is  detected but H$\beta$ is not. Values $\gtrsim 0.5$ may imply photoionization from the hard ionizing radiation field of the AGN or fast shocks ($\ge 300$ \kms). 
    \label{fig:bpt_map}}
\end{figure}

To determine the extent of the AGN influence, we take the ratio of \oiii\ 5007 \AA\ to H$\beta$, shown in Figure \ref{fig:bpt_map}.  We highlight areas where the outflow component is detected in \oiii\ but not H$\beta$ by the gray contours. The position angle (PA) of the radio jets \citep[160$^\circ$;][]{pedlar1990} within the central $\sim$20\arcsec\ is plotted as a dashed teal line where the southwest jet is coming towards the line of sight and the northeast jet is moving away from the line of sight. There is an enhancement in the host galaxy \oiii\ / H$\beta$ ratio along the PA of the receding jet, which may imply that the jet and/or AGN ionization cone is interacting with and injecting energy into the host galaxy. Simulations show that both shocks and AGN photoionization can provide the hard ionizing radiation field required for enhanced \oiii\ / H$\beta$ \citep{moy2002, rich2011}. The outflow \oiii\ / H$\beta$ ratio is enhanced compared to the host galaxy, also implying AGN influence. The enhancement is largely uniform across the outflow, although the spaxels with solely redshifted outflow display slightly lower values. 

\begin{figure*}
    \centering
    \includegraphics[width=0.8\textwidth]{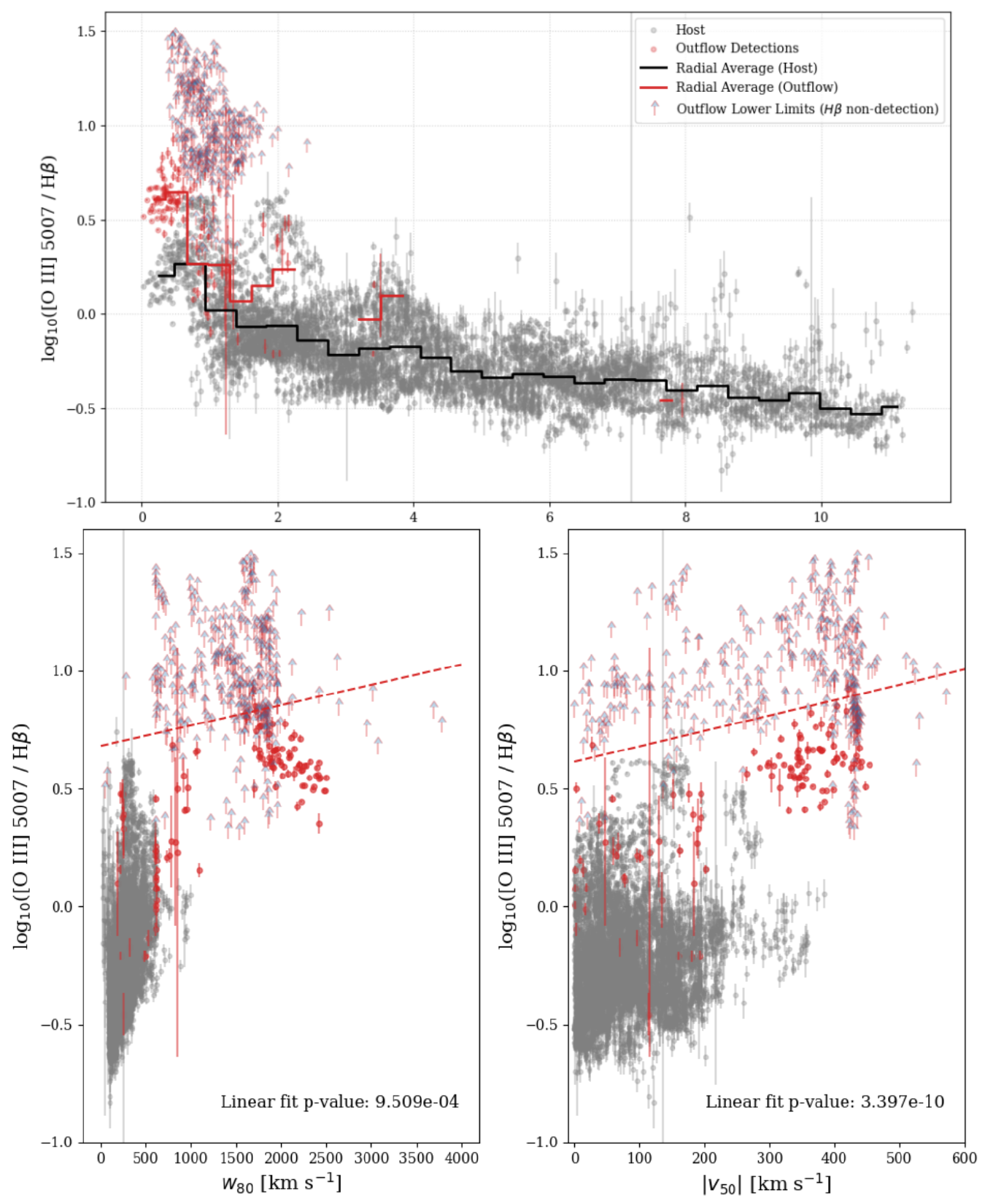}
    \caption{Top: log$_{10}$(\oiii\ / H$\beta$) plotted vs. distance from the center. Gray points show the host component, while red points represent the outflow component and purple lower limits mark outflow spaxels where \oiii\ is significantly detected but H$\beta$ is not. Red and black lines show the running means for the outflow and host components, respectively. The \oiii\ / H$\beta$ ratio declines with radius, showing that the influence of the AGN is limited to the central $\sim2$ kpc. 
    Bottom left: log$_{10}$(\oiii\ / H$\beta$) plotted vs. $w_{80}$ with points the same as the top panel.  Bottom right: log$_{10}$(\oiii\ / H$\beta$) plotted vs. $v_{50}$ with points the same as the top panel. In the bottom panels, the red dashed lines display linear fits to the outflow components (including lower limits; p-values shown in the lower right of each panel), showing that the outflow \oiii\ / H$\beta$ ratio is positively correlated with $w_{80}$ and $v_{50}$. 
    \label{fig:bpt_radial}}
\end{figure*}

In Figure \ref{fig:bpt_radial}, we plot the \oiii\ / H$\beta$ ratio vs distance from the center, $w_{80}$, and $v_{50}$, tests which have been used to investigate the source of the ionization \citep{veilleux1995}. Lower limits are included for spaxels where an \oiii\ outflow component is detected but not H$\beta$. We estimated the H$\beta$ flux upper limit assuming a Gaussian with the same width as \oiii\ and amplitude 3$\times$ the standard deviation around H$\beta$.
Both the host and outflow components exhibit peak ionization ratios in the nucleus that decline with distance, suggesting that the primary influence of the AGN is concentrated within the central $\sim2$ kpc.
In the outflow ratio, we see an increase in both $w_{80}$ and $|v_{50}|$ with increasing \oiii\ / H$\beta$, as shown by the bottom panels of Figure \ref{fig:bpt_radial}. 
The high values of $w_{80}$ and $|v_{50}|$ are further evidence of AGN influence. 

\begin{figure}
    \centering
    \includegraphics[width=\columnwidth]{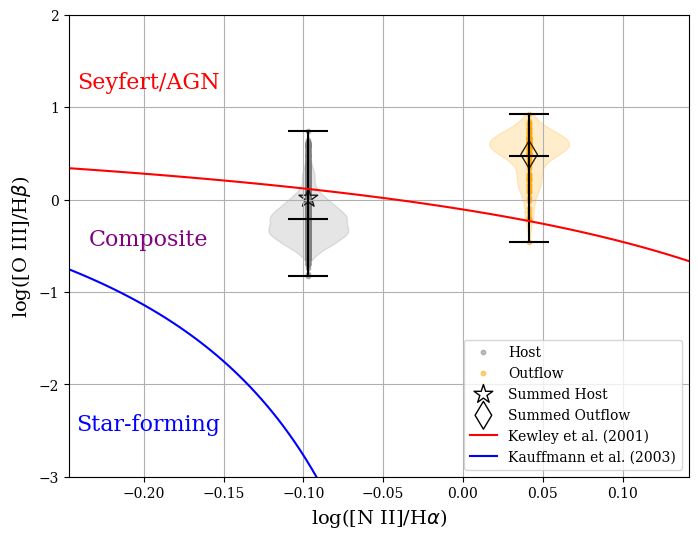}
    \caption{BPT diagram with violin plots of log$_{10}$(\oiii\ / H$\beta$). Gray points show host galaxy values and yellow points show outflow values. The log$_{10}$(\nii\ / H$\alpha$) values are estimated from Figure 3 in \citet{Gendron-Marsolais2018} where the outflow points use the log$_{10}$(\nii\ / H$\alpha$) value close to the nucleus and the host galaxy points use the log$_{10}$(\nii\ / H$\alpha$) value around 5-10 kpc from the center. Blue and red lines are the \citet{Kauffmann2003} empirical star-forming galaxy and \citet{kewley2001} maximum starburst lines, respectively. 
    \label{fig:bpt_violin}}
\end{figure}

We further contextualize these results in Figure \ref{fig:bpt_violin} by placing violin plots of the \oiii\ / H$\beta$ ratio on a classic Baldwin, Phillips \& Terlevich (BPT) plot \citep[][]{baldwin1981, veilleux1987}, where the \nii\ / H$\alpha$ ratios are estimated from Figure 3 in \citet{Gendron-Marsolais2018}. Close to the nucleus, the outflow component dominates the flux, so we utilize the \nii\ / H$\alpha$ value closest to the center for the outflow points. The host galaxy points use the \nii\ / H$\alpha$ value 5-10 kpc from the center, where the outflow is not detected. Based on the error bars in \citet{Gendron-Marsolais2018}, we estimate an uncertainty of $\pm 0.3$ in linear \nii\ / H$\alpha$. We overplot diagnostic lines from \citet{Kauffmann2003} in blue and \citet{kewley2001} in red. 

Although the \nii\ / H$\alpha$ ratio is imprecise, the strong \oiii\ / H$\beta$ ratio (log$_{10}$(\oiii\ / H$\beta$) $\le 0.93$) places the outflow firmly within the AGN/Seyfert regime, similar to enhanced ratios seen in other AGN outflows \citep[e.g.,][]{kader2026}. The approximate position of the outflow points on the BPT diagram is also consistent with fast shock models  \citep[$300-600$ \kms;][]{dopita1995, allen2008}. In the future, additional line tests (such as \oii\ / \oiii, \sii\ / H$\alpha$, and \oi\ / H$\alpha$ vs line width) could help determine between AGN photoionization and shocks as the ionizing source \citep{moy2002}. 
The host galaxy component has a slightly lower ratio at ($-0.83<$ log$_{10}$(\oiii\ / H$\beta$)$<0.75$) with the higher values lying along the jet PA and around the central 1 kpc implying AGN influence, while the rest of the host galaxy lies in the AGN-starforming composite regime.  

\subsection{Mass and Energetics}
We use the outflow component luminosities of H$\beta$ and \oiii\ 5007 \AA\ to estimate the warm ionized outflowing mass, following the methods in \citet{carniani2015}, detailed in \citet{veilleux2020}. There are four main assumptions: 
\begin{enumerate}
    \item $n_\mathrm{H}=10\times n_\mathrm{He}$ \citep[i.e., assuming a solar elemental abundance ratio;][]{Asplund2009};
    \item An oxygen abundance of solar [O/H] = $-$3.21\citep{Nicholls2017};
    \item Line emissivities calculated from \pyneb\ \citep{Luridiana2015} (assuming constant $T = 10^4$ K, which is appropriate for AGN or starburst photoionized gas), and constant $n_e = 150$ cm$^{-3}$, which was estimated as the median value of $n_e$ calculated from a \sii\ ratio map of the central 2 kpc via Vigneron et al. (in prep.);
    \item Electron densities lie below their critical densities so that collisional de-excitation is unimportant.
\end{enumerate}
This results in
\begin{equation}
\label{eqn:massneii}
\begin{split}
    M_{out, ion}^{\mathrm{\oiii}} = {} & 2.7 \times 10^8 \left(\frac{C}{10^{[O/H]-[O/H]_\odot}}\right) \\
    & \times \left(\frac{L_{\oiii}}{10^{44}\mathrm{erg/s}}\right) \left(\frac{\langle n_e \rangle}{150 \mathrm{\ cm}^{-3}}\right)^{-1} M_{\odot}
\end{split}
\end{equation}
\begin{equation}
\label{eqn:massneiii}
    M_{out, ion}^{\mathrm{H\beta}} = 3.6 \times 10^9 C \left(\frac{L_{H\beta}}{10^{44}\mathrm{erg/s}}\right)\left(\frac{\langle n_e \rangle}{150 \mathrm{\ cm}^{-3}}\right)^{-1} M_{\odot},
\end{equation}
where $C \equiv \langle n_{e} \rangle^2/\langle n_{e}^2 \rangle $ is the electron density clumping factor, which we assume is of order unity, $L$ is the luminosity, and $\langle n_{e} \rangle$ is the average electron density. We derive masses of $M_{out, ion}^{\mathrm{[OIII]}} = (8.5 \pm 0.8)\times 10^{5}$ M$_\odot$ and $M_{out, ion}^{\mathrm{H\beta}} = (2.7 \pm 0.3)\times 10^{6}$ M$_\odot$, where the uncertainties take into account the flux error as well as an approximate error on $n_e$ of $\pm100$ cm$^{-3}$. The H$\beta$-derived mass is a factor of $\sim$3 larger than the \oiii-derived mass, similar to \citet{carniani2015}, where they attribute the discrepancy to the volume of H$\beta$-emitting gas being larger than that of \oiii-emitting gas and infer that the \oiii-derived masses should be taken as lower limits. 

With an estimated escape velocity $v_{\mathrm{esc}} \approx 1500$ km s$^{-1}$ for NGC 1275 \citep{riffel2020}, we integrate over the outflow line profiles and find that, respectively, $\sim$ 10\% and 20\% of the \oiii\ and H$\beta$ outflow component line fluxes are emitted by gas with projected velocities in excess of the escape velocity. Assuming the warm-ionized gas mass directly scales with line flux, this implies that $10-20$\% of the outflowing ionized material will escape the galaxy and end up in the intracluster medium, while the other $80-90$\% will fall back and remain available for further star formation. 

From the outflow mass, we derive the mass outflow rate
\begin{equation}
\label{eqn:massoutflow}
    \dot M_\mathrm{out} = \frac{M_\mathrm{out} v_{\mathrm{out}}}{R_\mathrm{out}},
\end{equation}
momentum outflow rate, 
\begin{equation}
\label{eqn:momentumoutflow}
    \dot p_\mathrm{out} = \dot M_\mathrm{out} v_{\mathrm{out}},
\end{equation}
and outflow power, 
\begin{equation}
\label{eqn:energyoutflow}
    \dot E_\mathrm{out} = \frac{1}{2} \dot M_\mathrm{out} v_{\mathrm{out}}^2,
\end{equation}
where $v_{\mathrm{out}}$ is the median outflow velocity of each spaxel ($v_{50}$), and $R_{\mathrm{out}}$ is the radial distance from the center of each spaxel. For \oiii, we derive $\dot M_\mathrm{out} = 1.7 \pm 0.4$ M$_\odot$ yr$^{-1}$, $\dot p_\mathrm{out} = (3.9\pm 0.9)\times 10^{33}$ dyne, and $\dot E_\mathrm{out} = (7.2\pm 1.5)\times 10^{40}$ erg s$^{-1}$. For H$\beta$, we derive $\dot M_\mathrm{out} = 2.7 \pm 0.7$ M$_\odot$ yr$^{-1}$, $\dot p_\mathrm{out} = (3.3\pm 0.8)\times 10^{33}$ dyne, and $\dot E_\mathrm{out} = (4.0\pm 0.8)\times 10^{40}$ erg s$^{-1}$. 

\subsection{Comparison to Previous Observations}

We can compare our mass outflow rate to the various inflow and outflow rates from the literature. \citet{riffel2020} detect outflows in \feii\ 1.2570 $\mu$m and H$_2$ 1$-$0 S(1), probing neutral hydrogen and hot molecular gas phases, respectively. Their median velocities ($150-200$ \kms) and line widths ($w_{80}$ up to 2000 \kms) are smaller than the kpc-scale ionized outflow we observe. They find a comparable outflow mass ($6.5\times 10^5$ M$_\odot$) and mass outflow rate (1.6 M$_\odot$ yr$^{-1}$), although they use a higher electron density ($n_e = 500$ cm$^{-3}$, assuming a typical value for AGN), which deflates this rate compared to ours by a factor of $2-4$. As \feii\ probes neutral hydrogen due to its lower ionization potential of 7.9 eV, we are seeing a trend similar to other AGN-driven outflows where the warm ionized phase often contains the fastest moving gas. 

\citet{nagai2019} detect unresolved blueshifted absorption in HCN(3$-$2) and HCO+(3$-$2), which they interpret as cold molecular outflowing clumps with an estimated mass of 3 M$_\odot$ at a radius of $\sim$0.1 pc from the nucleus. Using Equation \ref{eqn:massoutflow} with their upper bound on the cloud velocity of 600 \kms, we derive a cold molecular mass outflow rate of $\sim$0.02 M$_\odot$ yr$^{-1}$, two orders of magnitude smaller than the warm ionized outflow rate and about the same as the hot molecular outflow rate from \citet{riffel2020}. However, since the cold molecular outflow is not resolved and may be outshone by the strong cold molecular disk component, further investigation is required to obtain a more accurate cold molecular mass outflow rate.

\citet{nagai2019} also estimate an average accretion rate of the cold molecular gas disk within a radius of 100 pc as 1$-$10 M$_\odot$ yr$^{-1}$ with an accretion timescale of $4\times 10^7$ yr, assuming a turbulent velocity dispersion of 25 \kms.
Our derived warm ionized mass outflow rate is on par with the estimated cold molecular gas disk accretion rate.
Assuming a maximum outflow radius of 2.5 kpc and average outflow $v_{50}$ of 300 \kms\ (see Figure \ref{fig:pv}), we estimate a warm ionized outflow timescale of $8\times 10^6$ yr, only a factor of 5 shorter than the accretion time scale that could be reconciled by a slightly smaller disk or one with a larger turbulent velocity. 
We could be observing direct evidence of how mechanical AGN feedback converts the inner pc-scale cold molecular accretion disk flow into a kpc-scale warm ionized wind, heating and expelling gas that would have otherwise been material for star formation. 

It is also worth noting that the jet-inflated cavity power is 218$^{+19}_{-26} \times 10^{42}$ erg s$^{-1}$ for the northern Perseus bubble and 289$^{+37}_{-80}\times 10^{42}$ erg s$^{-1}$ for the southern bubble \citep{timmerman2022}, both 3$-$4 orders of magnitude larger than our ionized outflow $\dot{E}_{out}$. The energy injection from the outflow is minimal compared to the radio bubbles, although the outflow probes timescales that are about an order of magnitude shorter than the cluster cooling time of $10^8$ yr \citep{fabian2003}.

\subsection{Origin of the Outflow}

To investigate the source of the energy for the outflow, we investigate both energy-conserving and momentum-conserving scenarios. In the energy-driven case, the wind cools inefficiently and expands as a hot bubble, transferring up to $\sim$5\% of the AGN bolometric luminosity to the outflow, while in a momentum-conserving flow driven by radiation pressure, $\dot p_{out} \simeq L_{Edd}/c$. We find that outflow’s kinetic power corresponds to 0.02\% of the AGN radiation power \citep[$L_{AGN}=10^{44.55}$erg s$^{-1}$][]{koss2022}, and the outflow momentum rate $\dot p_{out} \simeq 0.4 L_{Edd}/c$, implying a momentum-driven flow with relatively weak coupling efficiency. 

\subsection{Jet Powered Outflow?}

The powerful jets \citep[$P_{jet} = (1.5^{+1.0}_{-0.3})\times 10^{44}$ erg s$^{-1}$;][]{rafferty2006} in this system could also provide some or all of the outflow momentum. Here we discuss two different jet-driven outflow scenarios. 

For $P_{jet}\sim 10^{44}$ and hot-phase number density of 0.1, \citet{Wagner2012} predicts the maximum mean radial velocity of accelerated clouds to be $\sim 400$ \kms, while diffuse ablated material is accelerated to several 1000 \kms. Our projected velocities ($|v_{50}| \le 570$ \kms) lie within these two ranges, perhaps implying a mixed composition of both dense clouds and diffuse material. At the estimated timescale of our wind ($10^6-10^7$ yr), \citet{Wagner2012} predicts a high fraction of wind kinetic energy to injected jet power ($\sim 0.3$). Assuming constant energy injection, our outflow has a much smaller ratio of $\sim 0.0005$. 

For $P_{jet}\sim 10^{44}$ and central density of warm clouds $\sim 150$ cm$^{-3}$, \citet{Mukherjee2016} predicts a mass-weighted mean radial velocity of 300 \kms\ at $\sim 4\times10^6$ yr, which is a bit lower than our flux-weighted mean projected velocity of $-440$ \kms; however, the simulation can easily reach 400 \kms\ at a higher jet power of $P_{jet}\sim 10^{45}$. Similarly to \citet{Wagner2012}, \citet{Mukherjee2016} also predicts a fraction of wind kinetic energy to injected jet power ($\sim 0.15$) much higher than our estimated ratio of $\sim 0.0005$. However, since both the central AGN and radio jets in NGC 1275 are variable \citep{fabian2015,Foschi2025} and the fact that the outflow opening angle appears to not follow the jet axis \citep[see also VV 340a;][]{kader2026}, the assumption of constant and direct energy injection by the jet is likely an oversimplification and thus the injected jet power should likely be much lower. Future work should focus on establishing better geometric models of the outflow and jets in order to better estimate the injected jet power. 
 
\section{Conclusion}
\label{sec:conclusion}

We analyzed the KCWI IFS data of the central $\sim 10$ kpc of NGC 1275. These new data resolve the extent of the warm ionized outflow to kpc scales for the very first time. We used \badass, a Bayesian analysis algorithm, to fit the complex emission line profiles, decomposing the outflow from the host galaxy emission. Our spatially resolved analysis of the central few kpc adds to the broad literature on this rich system, helping us to better understand the nature of feedback in cool core cluster radio galaxies. The main results of our analysis include: 

\begin{enumerate}    
    \item NGC 1275 shows evidence of a warm ionized gas outflow with \oiii-derived projected $v_{50}$ and $w_{80}$ up to $-570$ and 3780 \kms, respectively.  
    
    \item We derive an outflowing mass from H$\beta$ of $(2.7 \pm 0.3)\times 10^6$ M$_\odot$ and from \oiii\ of $(8.5 \pm 0.8)\times 10^5$ M$_\odot$. From the H$\beta$ mass, we derive a mass outflow rate of $2.7 \pm 0.7$ M$_{\odot}$ yr$^{-1}$, momentum outflow rate of $(3.3 \pm 0.8) \times 10^{33}$ dyne and energy outflow rate of $(4.0\pm 0.8)\times 10^{40}$ erg s$^{-1}$. From the \oiii\ mass, we derive a mass outflow rate of $1.7 \pm 0.4$ M$_{\odot}$ yr$^{-1}$, momentum outflow rate of $(3.9 \pm 0.9) \times 10^{33}$ dyne and energy outflow rate of $(7.2\pm 1.5)\times 10^{40}$ erg s$^{-1}$. 

    \item We find that the mass outflow rate is comparable to the cold molecular disk accretion rate derived by \citet{nagai2019}, which could imply a relatively balanced state between the pc-scale cold gas feeding the AGN and the warm ionized wind accelerated back out to kpc scales. 

    \item NGC 1275 displays an enhanced \oiii\ / H$\beta$ ratio along the receding jet position angle in the host galaxy, potentially identifying an ionization cone where the AGN ionizing photons preferentially escape. This enhancement along the jet PA is not seen in the outflow component.  
    
    \item We present a BPT diagram using \nii\ / H$\alpha$ ratio values estimated from \citet{Gendron-Marsolais2018}. The outflow points show clear signs of AGN influence and may be consistent with fast shock models \citep[$300-600$ \kms;][]{allen2008}. Excluding the spaxels along the jet axis showing enhanced \oiii\ / H$\beta$, the host galaxy points mostly lie in the AGN-starforming composite range, implying both star formation and AGN influence.

\end{enumerate}

\begin{acknowledgments}
The authors wish to recognize and acknowledge the very significant cultural role and reverence that the summit of Maunakea has always had within the Native Hawaiian community. We are most fortunate to have the opportunity to conduct observations from this mountain.

This research has made use of the Keck Observatory Archive (KOA), which is operated by the W. M. Keck Observatory and the NASA Exoplanet Science Institute (NExScI), under contract with the National Aeronautics and Space Administration.

K.Y.D. acknowledges partial financial support by the U.S.-Japan Fulbright Program for this research. H.I. is supported by the HIRAKU-Global Program, which is funded by MEXT’s ``Strategic Professional Development Program for Young Researchers." V.U. and J.K. gratefully acknowledge partial funding support from NASA ADSPS grant \#80NSSC25K0169, National Science Foundation (NSF) Astronomy and Astrophysics Research Grant \#2536603, as well as STScI grant \#JWST-GO-08391.001-A, which was provided by NASA through a grant from the Space Telescope Science Institute, which is operated by the Association of Universities for Research in Astronomy, Inc., under NASA contract. T.G. acknowledges support from the Australian Research Council through Discovery Project DP210101945, funded by the Australian Government. 

We thank Benjamin Vigneron and his collaborators for providing the \sii\ ratio map from we which we derived the $n_e$ for our mass calculations. We also thank Marie-Lou Gendron-Marsolais and her collaborators for providing their data of NGC 1275, which aided in the early exploration of our data. 

\end{acknowledgments}

\begin{contribution}
K.Y.D. performed the analysis and wrote the manuscript. 
H.I. obtained the data, supervised the analysis, contributed to the interpretation of the results, and edited the manuscript. 
T.B. assisted with the analysis and edited the manuscript.
S.V. supervised the analysis, contributed to the interpretation of the results, and edited the manuscript.
V.U. obtained the data, contributed to the interpretation of the results, and edited the manuscript. 
T.G. obtained the data and edited the manuscript. 
J.K. assisted with data reduction and edited the manuscript. 
R.M. obtained the data.

\end{contribution}

\facility{Keck:II \citep[KCWI;][]{kcwi2018}}

\software{\badass\ \citep{sexton2021},
\astropy\ (\citealt{astropy:2013}, \citealt{astropy:2018}, \citealt{astropy:2022}), 
\matplotlib\ \citep{matplotlib2007},
\numpy\ \citep{numpy2020},
\scipy\ \citep{SciPy2020}, \pyneb\ \citep{Luridiana2015}
\photutils\ \citep{Bradley2026}.
We utilized GitHub CoPilot to assist in writing parts of the code for the analysis, accessed through the Visual Studio Code extension GitHub Copilot Chat (ver. 0.35.0). We utilized Gemini 3 Flash to help write code to create some of the figures as well as to reword a few sentences in the manuscript for clarity. 
}

\bibliography{sample701}{}
\bibliographystyle{aasjournalv7}

\end{document}